\input{aipcheck}

\documentclass[
    ,final            
  ]
  {aipproc}

\layoutstyle{8x11double}

\begin{document}

\title{Fundamental Parameters of Cepheids: Masses and Multiplicity }

\classification{90}
\keywords      {Cepheids, masses, multiplicity, evolutionary
tracks, star formation}

\author{Nancy Remage Evans}{
  address={SAO, 60 Garden St., MS 4, Cambridge MA 02138}
}

\begin{abstract} 
 Masses determined from classical Cepheids in binary 
systems are a primary test of both pulsation and 
evolutionary calculations.  The first step is to 
determine the orbit from ground-based radial 
velocities.  Complementary satellite data from Hubble, 
FUSE, IUE, and Chandra provide full information about
the system.  A summary of recent results on masses
is given. Cepheids have also provided copious information 
about the multiplicity of massive stars, as well as 
the distribution of mass ratios
and separations.  This provides some important 
constraints for star formation scenarios including differences
between high and low mass results and differences 
between close and wide binaries. 

\end{abstract}

\maketitle


\section{Introduction}

In addition to being  primary extragalactic 
distance calibrators, classical Cepheids provide an 
accurate luminosity for moderately massive stars evolved 
beyond the main sequence to the core He burning phase.  For
this reason, when a mass can be measured, they provide a 
benchmark to test evolutionary calculations.  As was appreciated more
than 4 decades ago the masses inferred from hydrodynamic pulsation  
calculations disagreed with the evolutionary track predictions,
though the discrepancy (``the Cepheid mass problem") has been
greatly reduced with revised opacity values.  

In a related topic, Cepheids have provided us with a wealth of 
information about the binary/multiple properties  
of intermediate mass stars, which are important to
check against the predictions of star formation calculations.  
Among these diagnostics are 
the fraction of binary or multiple systems, the
distribution of mass ratios, the maximum separation of systems, and the
difference in any of these between massive systems and 
low mass systems.  

\section{Binary Parameters }

\subsection{Mass Ratios }

One area where Cepheid systems provide particularly well defined
characteristics
is the mass ratios.  Because Cepheids are cool evolved stars,
when they have hot companions (late B or A spectral type), the
companions completely dominate at ultraviolet wavelengths.  
A survey of 76 of the brightest Cepheids (Evans 1992)
with the International Ultraviolet Explorer (IUE) satellite
 identified
{\it all} such hot companions.  Because the spectral region
1200 to 2000 \AA\/ is very temperature sensitive in this spectral
range, a very acurate temperature or spectral type can be determined 
from these spectra.  A standard zero-age main sequence (ZAMS) 
relation provides an accurate mass of the secondary (M$_2$)
from these spectral types.  
A  theoretical Cepheid mass-luminosity relation  provides the 
mass of the primary (M$_1$).  
Fig. 1 shows the distribution of mass  ratios determined
in this way.  In Fig. 1, the Cepheid mass 
assumes no main sequence core convective overshoot (as 
discussed in Evans, 1995).   This is an upper limit to
the predicted mass of the Cepheid, and it is likely that they
are somewhat smaller.  This should not change the shape of the
distribution, however. It only means that the mass ratios are 
a little underestimated.

\begin{figure}
  \includegraphics[angle=0,width=60mm]{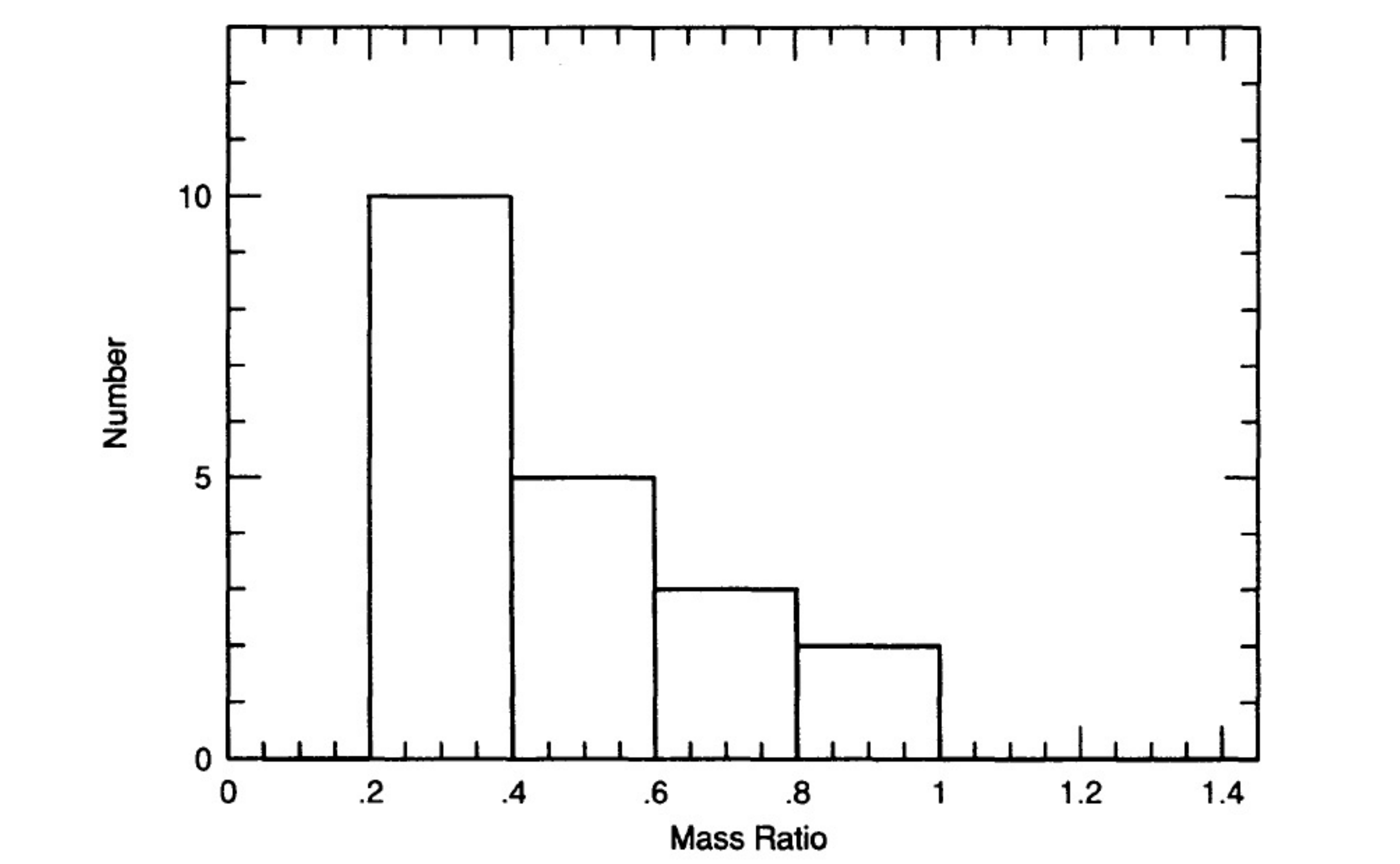}
  \caption{The mass ratio M$_2$/M$_1$ frequency histogram.    
The Cepheid mass (M$_1$) assumes no main sequence core 
convective overshoot, which is an upper limit to the primary
mass.   Taken from Evans, 1995.}
\end{figure}

The preference for low mass companions in Fig. 1 is clear. There 
is an important caveat to this diagram.  All Cepheids with known
orbits have periods longer than a year.  This makes sense 
since stars in closer orbits would have overflowed their 
Roche lobes (and presumably coalesced) in their previous 
evolution through the red giant phase.  Figure 1, however, 
provides a strong contrast in the distribution of mass 
ratios between these systems and 
systems with periods less that 40 days which have a 
preference for equal mass pairs (Tokovinin 2000).

\subsection{Selection Effects/ Completeness }

In discussions of binary parameters, it is always important 
to assess the completeness the of the information
available, in particular which companions would be detected.
Cepheids provide unusually complete 
information because the spectra of hot companions can be
observed uncontaminated by the spectrum of the primary 
(the Cepheid).  As discussed by Evans, et al. (2005), 
we now have a list of 18 Cepheids which have ground-based orbits
and an observation of the secondary spectrum with IUE.  
In addition 8 of these have been observed in high resolution
with either the Hubble Space Telescope (HST) 
or IUE, which allows a definitive 
answer to whether the secondary it itself a binary.  For this 
well-observed sample of 18 stars, 8 and possibly 9 (44\%,
possibly 50\%) are triple systems.  That is, there is a very high 
fraction of triple systems among the known binary systems.  

\subsection{Low Mass Companions }

What about low mass companions?  Do we have any information 
about them?  Because Cepheids are young stars, their 
companions must also be young.  Young stars of types late F, 
G, K, and M are prolific X-ray producers because their coronae
are much more active  than those of older field stars.  An
optical image of a field within 0.1 pc of a Cepheid (the commonly
accepted radius outside which binary systems are thought 
to be disrupted by the galactic star field) will show 
numerous red stars.  An X-ray image will typically indicate a very 
small number which are young, and hence possible physical 
companions.  We (Evans, Guinan, et al. 2009) have recently 
observed Polaris with the ACIS camera on the Chandra X-ray 
satellite.  This image showed that two cool stars suggested 
to be companions (Polaris C and D) are not young stars.  However, 
it identified one and possibly two stars which might be.
The recent detections of three Cepheids on XMM X-ray images 
(Engle et al. 2009) means that it is likely that the 
Cepheid contributes to the X-ray source at the location of 
Polaris A =Aa + Ab, although the spectroscopic companion Ab 
has a suitable spectral type (F6 V) to produce X-rays also.  

\section{Masses }

Several new results have been added to the list of Cepheid 
masses in recent years.  Currently masses have been determined 
in three ways.  In some cases the high resolution 
ultraviolet spectra of
the companion have provided an orbital amplitude for the 
secondary.  This can be combined with the orbital velocity
amplitude of the primary and the mass of the secondary 
from IUE spectra to obtain the mass of the Cepheid (primary).
Second, recently Benedict, et al. (2007) have observed astrometric
motion in Hubble Fine Guidance Sensor (FGS)
observations for two Cepheids.  
Combining this with the orbit of the Cepheid 
and the mass of the companion also produces a Cepheid mass. 
Finally, the 30 year orbit of Polaris has been resolved with 
the Hubble Advanced Camera for Surveys (ACS), and the 
companion has been detected.  This results in the first 
purely dynamical determination of the mass of a Cepheid. 

\subsection{S Mus }

We will now discuss three recent results on Cepheid masses.  The
Cepheid S Mus has the hottest known companion.  The high 
ultraviolet flux of S Mus B means that  it could be 
observed with the high resolution echelle mode of the HST  
Goddard High Resolution Spectrograph (GHRS, B\"ohm-Vitense,
et al. 1997), and hence it has the most accurately measured 
orbital velocity amplitude of a secondary.  Unfortunately, 
because of the high temperature of the companion there is a 
degeneracy between the energy distribution and the reddening 
in the IUE spectral region, 
and the temperature cannot be as well
determined as for cooler companions.  For this reason, we (Evans, 
et al. 2006) obtained a FUSE spectrum, and identified 
regions which are 
relatively free of H$_2$ absorption but particularly
temperature sensitive.  Comparison with appropriate standard
star spectra in these areas resulted in an improved 
temperature for the companion, and hence an improved 
Cepheid mass.  

\subsection{W Sgr }

\begin{figure}
  \includegraphics[angle=0,width=60mm]{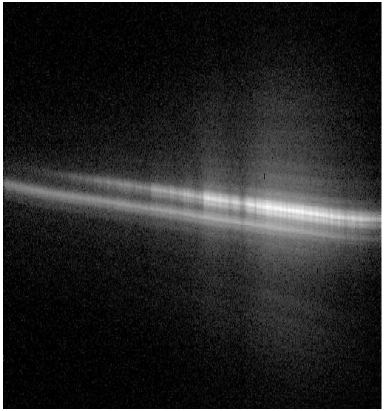}
  \caption{An HST STIS spectrum of the W Sgr system.  
Wavelength increases from left to right from 
approximately 1792 to 3382 \AA. The system is 
spatially resolved in  the approximately vertical 
direction.  The hottest component in system is the 
lower one.  The upper spectrum is the combined Cepheid
plus the cooler companion, the component in the 
spectroscopic binary.  
   Taken from Evans, et al. 2009.}
\end{figure}

W Sgr is a system for which Benedict, et al. (2007) measured 
astrometric motion.  It has a small orbital velocity 
amplitude, and a spectral type for the companion W Sgr B 
of A0 V from IUE spectra.  Fig. 2 shows a Hubble Space
Telescope Imaging Spectrograph (STIS) spectrum of the system
(Evans, et al. 2009) which changed our understanding completely.  
The spectroscopic binary Aa + Ab 
would not be resolved in the image, and is the top spectrum  
with the Cepheid dominating the spectrum at the long 
wavelength (right) side. The hottest star (which would 
dominate the IUE spectrum from 1200 to 2000 \AA) is clearly
resolved as the third star in the system.  The 
spectroscopic binary (upper spectrum) was carefully 
extracted.  Comparison with supergiant spectra comparable 
in temperature to the Cepheid Aa showed that no flux from the
companion Ab was identified.  An upper limit to the mass 
(spectral type) of the companion Ab was obtained, resulting
in an upper limit to the Cepheid mass.  

\subsection{Polaris }

Finally, Polaris is a low amplitude Cepheid which has long 
been known to be a member of a 30 year spectroscopic 
binary system (Kamper 1996).  Recently, Wielen et al. (2000)
used the proper motion of Polaris from {\it Hipparcos}
observations to determine the inclination of the system. 
As the final piece of the puzzle we (Evans, et al. 2008) 
used the High Resolution Channel (HRC) of the Advanced 
Camera for Surveys (ACS) on Hubble to detect the faint 
companion and hence measure the separation between the 
primary and the secondary (Fig. 3).  A second observation a year
later detected orbital motion of the companion.   
The right hand image in Fig. 3 shows that the image of 
Polaris B taken at the during the 2006 observations, which
was used to determine the point spread function (PSF) of 
the Polaris A image.  Note that the PSF of Polaris B 
has an asymmetry similar to that of Polaris A, but with 
no sign of an artifact at the location of Ab in the left
and middle images, confirming the detection of Polaris Aa.  

Combining all these data we (Evans, et al. 2008) determined a
mass of the Cepheid, Polaris Aa of  $4.5^{+2.2}_{-1.4} \, M_\odot$,
the first purely dynamical mass of a Cepheid. 

We also determined the properties of the close companion Ab.  
It is 5.4 mag fainter than the Cepheid at 2255 \AA, with
an inferred spectral type of F6 V.  This is consistent with
the mass $1.26^{+0.14}_{-0.07} \, M_\odot$ found from the 
mass solution.  

Polaris B (18'' from the Cepheid) has a flux at 2255 \AA\/
consistent with a spectral type of F3 V -- F4 V at the 
distance of Polaris.  It also has a proper motion very close
to that of Aa + Ab, consistent with orbital motion in a long-period
bound system.

\begin{figure}
\includegraphics[width=150mm]{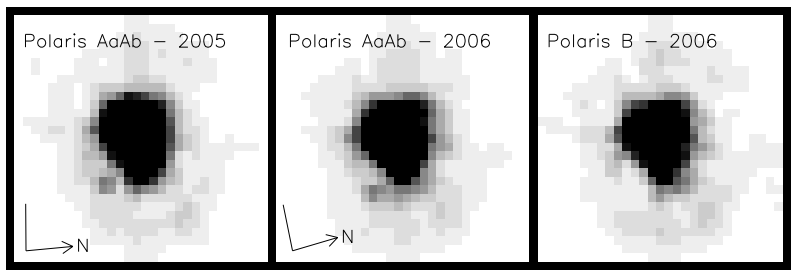}
  \caption{Co-added ACS HRC images of Polaris Aa (primary, Cepheid)
    and Ab (companion) taken in 2005 (left) and 2006 (middle) in 
0.85'' x 0.85'' images at 2255 \AA.  Ab is at the ``7 o'clock'' position. 
The right image shows the distant companion Polaris B taken during
the 2006 observations.  
Taken from Evans et al. 2008.}
\end{figure}

\section{Future Work }

We can expect improvements from future work, 
particularly in two ways.  
First, we should be able to improve the error 
bars on the masses of several of the systems. 
For Polaris, for instance, continued measurement 
of the motion of the companion Ab will improve
the mass determination.   
  In addition, future X-ray observations should improve our 
knowledge of the occurrence of low mass companions.


\begin{theacknowledgments}

Acknowledgments:  Many people contributed to 
these projects including  Ed Guinan, Scott Engle, Howard
Bond, Gail Schaefer, Derek Massa, Charles Proffitt, 
Scott Wolk, Massimo Marengo, Margarita Karovska, Ken
Carpenter, Erika B\"ohm-Vitense, and Giuseppe Bono. 
Funding for this work was provided by 
Chandra X-ray Center NASA Contract NAS8-39073
and STScI grant GO-10891.  

\end{theacknowledgments}



\begin{thebibliography}{9}

\bibitem{Benedict et al. 2007}
Benedict, G. F., McArthur, B. E., Feast, M. W., Barnes, T. G.,
Harrison, T. E., Patterson, R. J., Menzies, J. W., Bean, J. L.,
Freedman, W. L. 2007, AJ, 133, 1810


\bibitem{Bohm-Vitense et al. 1997}
B\"ohm-Vitense, E., Evans, N. R., Carpenter, K., 
Beck-Winchatz, B., and Robinson, R. 1997, ApJ, 477, 916



\bibitem{Engle et al. 2009}
Engle, S. G., Guinan, E. F., DePasquale, J., and Evans, N. 2009,
in the
proceedings of "Future Directions in Ultraviolet Spectroscopy'', in press

\bibitem{Evans 1992}Evans, N. R. 1992, ApJ, 384, 220


\bibitem{Evans 1995} Evans, N. R. 1995, ApJ, 445, 393.

\bibitem{Evans, et al. 2005}Evans, N. R., 
Carpenter, K. G., Robinson, R., Kienzle, F.,
and Dekas, A. E. 2005, AJ, 130, 789

\bibitem{Evans, et al. 2006}
Evans, N. R., Massa, D., Fullerton, A., Sonneborn, G., 
and Iping, R. 2006, ApJ, 647, 1387


\bibitem{Evans, et al. 2008}
Evans, N. R.,  Schaefer, G. H.,  Bond, H. E.,  Bono, G.,  
Karovska, M., Nelan, E.,     Sasselov, D., and   
Mason, B. D.  2008, AJ,
136, 1137

\bibitem{Evans, et al. 2009}
Evans, N. R., Massa, D., and Proffitt, C. 2009, 
AJ, 137,  3700

\bibitem{Evans, Guinan, et al. 2009}
Evans, N. R., Guinan, E., Engle, S., Wolk, S. J., 
Schlegel, E., Mason, B. D., and Karovska, M. 2009,
AJ, submitted

\bibitem{Kamper 1996}  Kamper, K. W. 1996, JRASC, 90, 140

\bibitem{Wielen et al 2000}
Wielen, R., Jahreiss, H., Dettbarn, C., Lenhardt, H., and Schwan,
H. 2000, A\&Ap, 360, 399


\end{thebibliography}
\end{document}